%% file: 0-main.tex
\documentclass[sigconf]{acmart}
\usepackage{booktabs} 
\usepackage[utf8]{inputenc}
\usepackage{url}
\usepackage{multirow}

\usepackage{graphicx}

\usepackage[ruled]{algorithm2e} 

\SetAlFnt{\small}
\SetAlCapFnt{\small}
\SetAlCapNameFnt{\small}
\SetAlCapHSkip{0pt}
\IncMargin{-\parindent}

\acmConference[HIP3ES-2019]{7th International Workshop on High Performance Energy Efficient Embedded Systems }{January 22, 2019}{Valencia, Spain}
\setcopyright{none}
\acmPrice{}
\acmISBN{}
\acmDOI{}
\acmYear{}

\begin{document}
\title{High performance scheduling of mixed-mode DAGs on heterogeneous multicores}

\author{Agnes Rohlin}
\affiliation{%
  \institution{Chalmers University of Technology}
}
\email{agnesrohlin@gmail.com}
\author{Henrik Fahlgren}
\affiliation{%
  \institution{Chalmers University of Technology}
}
\email{henrik\_fahlgren@hotmail.com}
\author{Miquel Peric\`as}
\affiliation{%
 \institution{Chalmers University of Technology}
}
\email{miquelp@chalmers.se}


\begin{abstract}


Many HPC applications can be expressed as mixed-mode computations, in which each node of a computational DAG is itself a parallel computation that can be molded at runtime to allocate different amounts of processing resources. 
At the same time, modern HPC systems are becoming increasingly heterogeneous to address the requirements of energy efficiency.
Effectively using heterogeneous devices is complex, requiring the developer to be aware of each DAG nodes' criticality, and the relative performance of the underlying heterogeneous resources. 

This paper studies how irregular mixed-mode parallel computations can be mapped on a single-ISA heterogeneous architecture with the goals of performance and portability.
To achieve high performance we analyze various schemes for heterogeneous scheduling, including both criticality-aware and performance-only schemes, and extend them with task molding to dynamically adjust the amount of resources used for each task. 
To achieve performance portability we track each DAG nodes' performance and construct an online model of the system and its performance. 
Using a HiKey960 big.LITTLE board as experimental system, the resulting scheduler implementations achieve large speed-ups when executing irregular DAGs compared to traditional random work stealing.

\end{abstract}

\maketitle

\input{1-intro.tex}

\input{2-back.tex}

\input{3-sched.tex}

\input{4-eval.tex}

\input{5-results.tex}

\input{6-conc.tex}

\bibliography{7-ref.bib}
\bibliographystyle{plain}

\end{document}

%% file: 1-intro.tex
\section{Introduction}

Modern HPC applications are often implemented as mixed-mode parallel computations, in which the nodes of a computational task DAG are themselves parallel computations that can be assigned varying amounts of processing resources~\cite{chakrabarti-jpdc97,wimmer-spaa11}. 
This model extends the classical single-threaded task-DAG model, in which each task can only be mapped to a single processor. 
The scheduling of such DAG problems has been heavily researched. For example, the greedy scheduling theorem --a classic result in scheduling theory-- states that, in the presence of abundant parallelism, keeping processors busy whenever there are ready tasks will result in linear speed-up~\cite{blumofe-jacm99}. 
As a result, many dynamic schedulers such as random work stealing follow the goal of maximizing resource usage while minimizing scheduling overheads.

However, the greedy scheduling theorem assumes that tasks executing in parallel do not interfere with each other. It also assumes that memory bandwidth is not a limiting factor. In practice, resources such as memory bandwidth, or shared cache capacity are heavily contended, leading to slowdown and non-linear scaling of greedy scheduling. One way to address this limitation is to convert the classical 1-task to 1-core scheduling problem into a hiearchical problem in which a global level schedules M-task-groups onto N-core-places, and a local level schedules the M-tasks onto the assigned N-cores. This scheme, known as Elastic Places, targets interference-free scheduling and allows to retain the greedy property in the global scheduler~\cite{Pericas2018}. Furthermore, this scheme maps naturally to mixed-mode parallelism when we consider each M-task-group to be a parallel computation in a mixed-mode task graph.

Elastic Places is implemented in a runtime library called XiTAO~\cite{Pericasweb} and has been shown to perform efficiently on homogeneous NUMA systems~\cite{Pericas2018}. 
To achieve higher efficiency, however, it is important to adapt XiTAO to heterogeneous platforms. As an additional limitation, XiTAO requires the programmer to determine the size of the N-core-places, which limits productivity and performance portability. 

This paper explores schemes to automatically determine resource partitions at runtime and researches how this knowledge can be used to exploit modern single-ISA heterogeneous platforms such as ARM's big.LITTLE. To this end we implement two heterogeneous schedulers inspired by CATS~\cite{Chronaki2015} and performance bias scheduling~\cite{Koufaty}, and we propose a performance trace table (PTT) to automatically learn the best partition sizes. 
We observe that this scheme not only keeps track of efficient parallelism levels, but it also is able to infer the load of the system, which allows the scheduler to target interference-free scheduling 
without the need to statically analyze the characteristics of the task DAG that is being executed. 
%
We evaluate the scheme using irregular DAGs composed of parallel subcomputations on a HiKey960 board with 4 big and 4 LITTLE ARMv8 cores. On DAGs with a parallelism degree ranging between 1.62 and 8.08, both CATS and performance bias outperform random work stealing, achieving up to 2.7$\times$ speed-up without any programmer involvement to select core partition sizes. 

Compared to prior art, this paper makes two contributions. First, it studies scheduling strategies for mixed-mode task graphs on heterogeneous hardware. Second, it explores how a performance tracer can be used to automatically select efficient hardware places (i.e.,~core partitions) for the execution of nested parallel computations.   


%% file: 2-back.tex
\section{Background}\label{chap:Bckg}

\subsection{Mixed-mode parallelism and XiTAO}\label{sec:xitao} 

\begin{figure}[tbh]
\centering
\includegraphics[width=0.75\columnwidth]{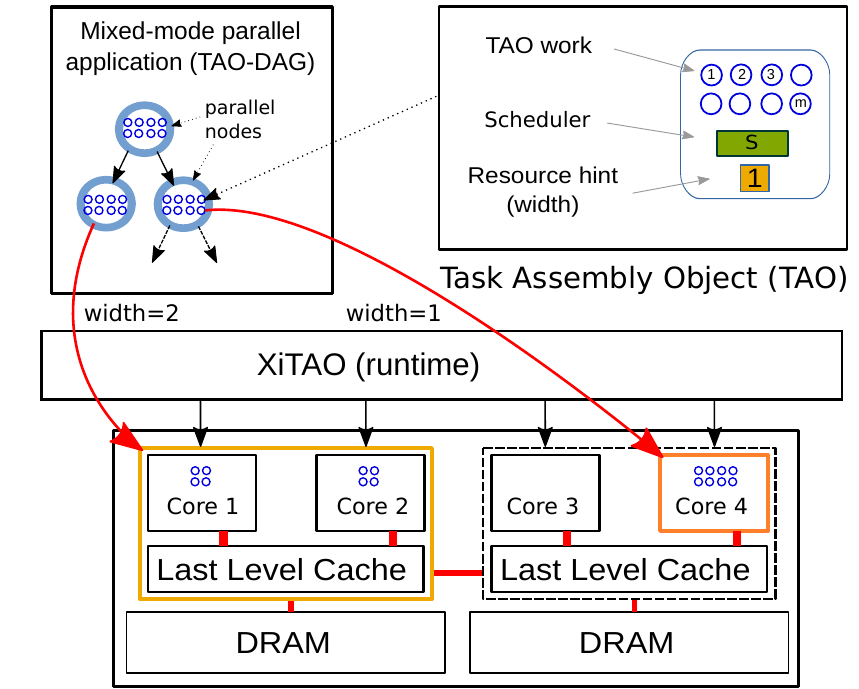}
 \caption{A mixed-mode parallel application running on a dual-socket dual-core system under the control of the XiTAO runtime.
 }
 \label{fig:xiTAO}
\end{figure}

Irregular applications are frequently implemented as mixed-mode parallel applications. In a mixed-mode application, a global task-DAG is composed of parallel nodes. The parallel nodes usually describe data parallelism, such as OpenMP {\tt for} loops, but other forms of parallel patterns, such as reductions are also common. In a mixed-mode parallel application, the parallel nodes can map the full system, or they can be scheduled concurrently to other tasks. The latter enables higher performance but its implementation raises several questions, such as: \textit{how to schedule parallel nodes to partially overlapping hardware allocations?} or \textit{how to minimize interference across parallel nodes?}    
Figure~\ref{fig:xiTAO} (top left) shows a sample DAG for a mixed-mode parallel application and its potential execution on a dual-socket dual-core system.

XiTAO is an execution model and runtime developed for resource-efficient and interference-free execution of mixed-mode parallel applications~\cite{Peric2015,Pericas2018,Pericasweb}. XiTAO targets multicore environments and is built on top of C++11's threading model. The basic idea of XiTAO is to minimize the overheads associated with fine-grained parallelism without sacrificing the performance gains by encapsulating sets of tasks into parallel subgraphs. This approach allows for interference-free scheduling by avoiding oversubscribing the resources of each subgraph. XiTAO does this by associating the parallel subgraph with an internal scheduler and a resource hint. These subgraphs, called \textit{Task Assembly Objects} (TAOs), are moldable entities allocated to a suitable hardware place as directed by its resource hint. 

The resource hint within a TAO allows for the runtime to treat TAOs as moldable and schedule them onto \textit{elastic places}. Elastic places are hardware places which are dynamically and asynchronously provided to TAOs at runtime. A hardware place refers to a particular collection of cores. In the context of XiTAO, the number of cores in a hardware place is called its {\it resource width}. A resource hint in a TAO refers to the width of a hardware place.
This scheme allows for locality-aware and interference-free scheduling as tasks can be bound within a TAO and scheduled onto adjacent cores. This strategy enables both temporal locality as well as spatial locality, and effectively exploits cache hierarchies in modern multicores. 
Note that the global XiTAO scheduler simply sees the TAOs as ''wider'' units of computation, i.e.~a black box filled with work. Since each TAO includes an embedded scheduler, TAOs support any type of work, from simple single-threaded tasks, to dependent set of tasks (DAGs) and even whole runtimes (e.g.~OpenMP, Intel's TBB). 

TAOs are scheduled in XiTAO via a global scheduling algorithm called Dynamic Place Allocation (DPA). In DPA, workers find ready TAOs using a method based on random work stealing. Upon selecting a ready TAO they allocate a resource partition using a centralized method. Finally, the worker threads execute the TAOs asynchronously. Overall, this results in a pipelined execution of TAOs with low overheads. The only centralized part of the scheme is the selection of resources. Fortunately, large places are usually not required for performance, thus the scheme overall has good scalability. More details on scalability can be found in~\cite{Pericas2018}.

A diagram of the XiTAO system, including the scheduler, an application DAG and the format of a TAO is shown in Figure~\ref{fig:xiTAO}. To avoid confusion with a potentially local work DAG, the global DAG (i.e.~the DAG of the mixed-mode application) is generally called the TAO-DAG in the XiTAO system. 

\subsection{Scheduling in a Heterogeneous Environment}\label{sec:hertrosched}
Task scheduling on a heterogeneous platform, contrary to a homogeneous platform, includes the problem of assigning the appropriate tasks to the most suitable cores. Most multicore scheduling approaches today assume equal performance. For example, dynamic scheduling techniques such as work-stealing or work-sharing do not consider the individual performance of cores.

Scheduling DAGs on heterogeneous multicores is a well studied problem in the context of single-threaded task DAGs~\cite{Topcuoglu2002,Cheng2010,Chronaki2015,Chronaki2017,Koufaty,VanCraeynest2012}. These schemes either assign a ranking to each tasks based on the critical path and then assign more critical tasks to faster cores~\cite{Topcuoglu2002,Cheng2010,Chronaki2015,Chronaki2017}, or they compute a best fit between tasks and cores and then schedule appropriately~\cite{Koufaty,VanCraeynest2012}. In this study we implemented the ideas behind the CATS~\cite{Chronaki2015} and Bias Scheduling~\cite{Koufaty} schemes. We describe them below along with HEFT~\cite{Topcuoglu2002}, a classical heterogeneous scheduler.

\subsubsection*{Heterogeneous Earliest-Finish-Time (HEFT)}
HEFT is a static scheduling method for heterogeneous task scheduling proposed by Topcuoglu et al.~\cite{Topcuoglu2002}. The HEFT algorithm consists of ranking the tasks of a DAG in order of longest path to finish and then assigning the highest-ranking tasks to the core that will minimize the overall finish time. An analysis of the DAG is done to calculate the execution time and communication cost of each node and edge before the tasks can be ranked. The tasks are then placed in a queue where the scheduler picks the top task and calculates which core will be able to finish this task earliest using insertion-based scheduling. 



 \subsubsection*{Criticality Aware Task Scheduling (CATS)}
 CATS is a dynamic scheduling approach where no prior knowledge about the execution time of the tasks is assumed~\cite{Chronaki2015,Chronaki2017}. 
 Instead, CATS solely uses the number of successors to find the critical path. The critical path is then put in a critical queue. Tasks from the critical queue are scheduled on high-performance cores and tasks from the non-critical queue are scheduled on lower-performance cores. 
 In~\cite{Chronaki2015}, Chronaki et al.~introduce the dynamic Heterogeneous Earliest Finish Time (dHEFT) algorithm as a reference to evaluate CATS. dHEFT uses the same principles as HEFT but instead of knowing the load of tasks prior to scheduling, discovers them at runtime. 

 \subsubsection*{Bias Scheduling in Heterogeneous Multicore Architectures}
 Bias Scheduling~\cite{Koufaty} is a proposed method for single-ISA heterogeneous multicore processors that tracks how different kinds of tasks perform on each core. The main idea is to categorize tasks into two groups: Tasks gaining large speedup by running on a big core compared to a LITTLE core and tasks gaining modest speedup by running on a big core. The speedup is approximated by accessing hardware counters for stall cycles. Tasks are then scheduled on big cores if they provide large speedup and on LITTLE cores if the speedup would be modest. 


%% file: 3-sched.tex
\section{Heterogeneous mixed-mode Scheduling}\label{chap:Imp}
This section explains the implementation of the heterogeneous, mixed-mode scheduler in XiTAO. We begin by describing the performance trace table and then describe the changes to the scheduling algorithm. 

\subsection{Performance Trace Table (PTT)} \label{sec:table}
To be able to dynamically affect the scheduling decisions based on the available resources, we implemented a \textit{performance} tracer of TAOs at runtime and a table to record task execution times. Although several scheduling implementations surveyed in Section~\ref{sec:hertrosched} assume prior knowledge of task loads~\cite{Topcuoglu2002,Cheng2010}, this is not applicable in our case where the runtime has no prior knowledge of the task execution times. A table has been implemented where for each TAO type, core and resource-width the execution time is recorded. In XiTAO, all TAOs are instances of a particular TAO class (type). This allows us to implement the performance tracer by instantiating one PTT for each TAO class.



\begin{figure}[tbh]
\centering
\includegraphics[width=0.5\columnwidth]{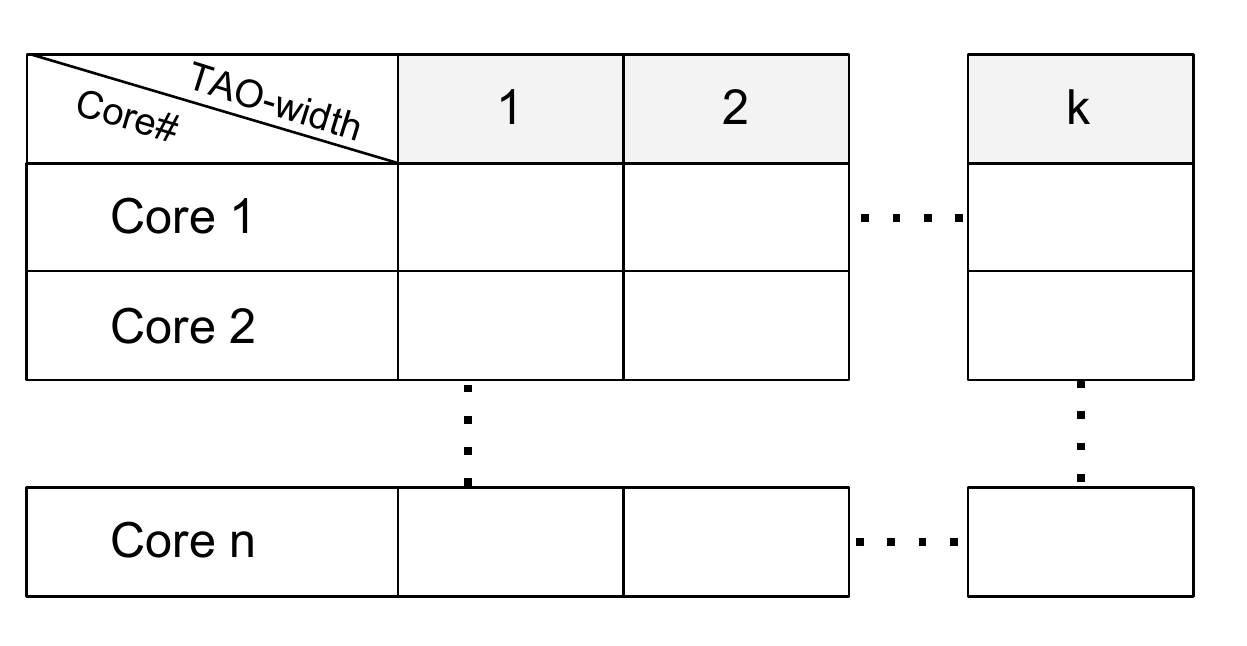}
\caption{The structure of the performance trace table (PTT) where n is the number of cores in our system and k is the maximum resource width, $log2(\#cores)$.}
\label{fig:time-table}
\end{figure}

The table is organized by $(core number)\times (resource width)$ as seen in Figure~\ref{fig:time-table}. The fields of the table are initialized to 0, which ensures that all configurations will be tested at runtime. Due to the distributed implementation of the scheduler, the table is organized to fit into cache lines where each core only accesses one cache line indexed with core number, hence avoiding false sharing. For each entry, the execution time is stored. 
The table is updated after each TAO execution with a weighted time of 1:4 to the old value of the table: $saved value=\frac{(4 * old value) + new value}{5}$. 

The table is updated always by the lowest ranked worker of a TAO to minimize cache migrations. This worker is called the \textit{leader} of the TAO. In XiTAO the leader of a TAO is decided when the TAO is distributed to the cores by the DPA. The leader is then set to $ \lfloor \frac{core}{width} \rfloor \times width $, where $core$ is the \textit{core} distributing the TAO and \textit{width} is the resource hint. Since the floor function is applied to the division, only a subset of cores are eligible to become leaders for large resource widths. 
For example, if core number \textit{seven} were to distribute a TAO with resource width \textit{four}, then core number \textit{four} would be chosen as leader. For the PTT this means that every core can have a recorded value of the TAO with $width=1$ but only every fourth core can have a recorded value of the TAO with $width=4$. By only allowing the leader to save its execution time, accesses to the table result in less sharers, but it can potentially skew the recorded result as the leader might not have the most representative view of the execution time of that TAO, i.e.~it could have the least or the most amount of work. Remember that workers enter and exit the execution of TAOs asynchronously~\cite{Pericas2018}. This is however dealt with as we are weighting the recorded values 1:4 and any particularly diverging values do not affect the table significantly. Although this results in an additional read of the table we found it more important to be resilient to divergent measurements as this table aids the scheduling decisions.

This implementation of tracing execution history requires no knowledge of the available resources as the cores simply update the corresponding index, independent of its resource type. This is beneficial not only for portability and potentially functional-heterogeneity, but we also expect it to be useful in the context of temporally added heterogeneity such as dynamic voltage frequency scaling (DVFS) caused by heat variations, or even interference caused by other TAOs and/or uncontrollable system activities such as background processes or interrupts.

\subsection{Heterogeneous Scheduling Extensions}\label{sec:scheduler}
We implemented two different scheduler extensions to XiTAO: criticality-based and weight-based scheduling. The first one targets the criticality of a DAG and the second aims to exploit the performance difference between the TAOs on the two core clusters of ARM's big.LITTLE architecture. The algorithms are inspired by the methods introduced in Section~\ref{sec:hertrosched}, especially Bias scheduling by Koufaty et al.~\cite{Koufaty} and CATS by Chronaki et al.~\cite{Chronaki2017}. Note that, contrary to the methods in Section~\ref{sec:hertrosched}, XiTAO is a distributed runtime without a central governing scheduler, thus we cannot implement identical solutions as CATS and Bias scheduling. Instead we focused on exploiting the benefits of XiTAO as we make use of the PTT to find suitable resources. 

The scheduling extensions were implemented within the mechanisms of the runtime, specifically in the commit-and-wakeup mechanism. The commit-and-wakeup is a routine within each TAO responsible for waking up depending tasks and is executed by the last core completing the execution of a TAO. Each call to the commit-and-wakeup checks which of the depending TAOs are ready for execution. The ready TAOs are then placed into the work-stealing queues or executed in-place, targeting locality. Making a scheduling decision within this mechanism allowed us to maintain the DPA and the notion of elastic places and resource allocation as it is, effectively preserving important key notions of the XiTAO runtime. None of the scheduling extension made any changes to the underlying load-balancing policy or to the DPA. 

\subsubsection{Criticality-based Scheduling}\label{sec:crit}
The CATS scheduling schema by Chronaki et al.~\cite{Chronaki2017} has been implemented as a pre-execution criticality-analysis where the critical path of the TAO-DAG to be executed is determined by the longest path and tasks are placed into two sets of centralized queues, critical and non-critical, depending on the criticality analysis. The decision of making the longest path the critical path is suitable to our runtime environment where no knowledge of task execution times is available prior to execution. Therefore, our criticality scheduling extension in XiTAO also shares this notion of critical path. 

\begin{figure}[tbh]
\centering
\includegraphics[width=0.35\columnwidth]{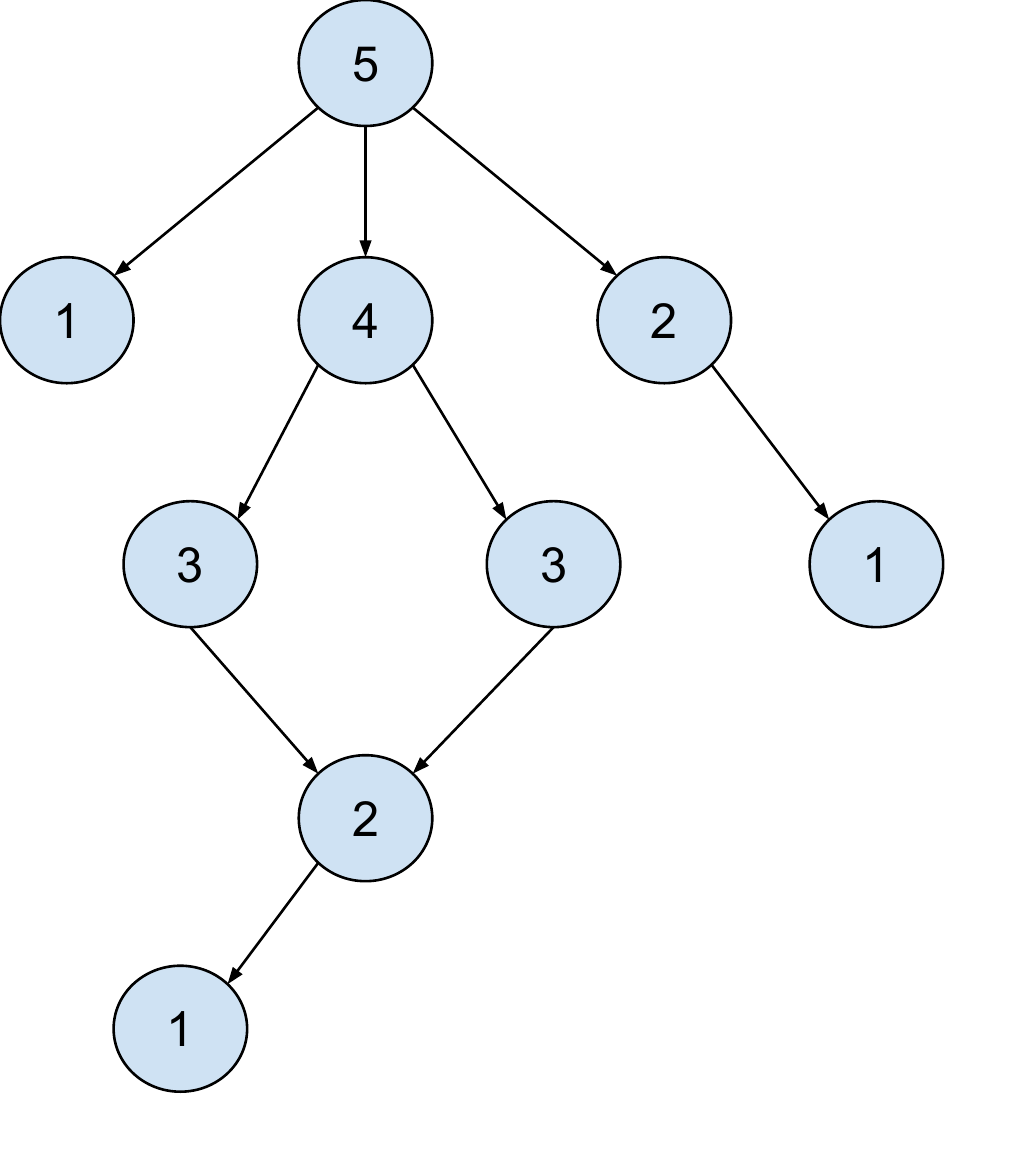}
\caption{An example of the criticality values assigned to a DAG for the criticality-based scheduling extensions.}
\label{fig:critassign}
\end{figure}

For our implementation, the criticality analysis of the TAO-DAG is done as the runtime is started by calling a recursive function which assigns criticality values on the pushed TAOs (i.e..~the ready TAOs) and their successors. The recursive function traverses top-down through the TAO-DAG until it reaches the end node(s) and assigns each node a criticality value of $max(crit(child))$, effectively resulting in the first node of the longest path having the highest criticality value. An example of the criticality assignment of a small DAG can be seen in Figure~\ref{fig:critassign} where the DAG has been traversed top to bottom recursively. 

The criticality values are then used whenever a TAO is woken up in the commit-and-wakeup mechanism. When a TAO is being woken up, its criticality value is compared to the maximum criticality TAO that is currently running and the TAO is then deemed either critical or non-critical in comparison. The runtime keeps track of the currently maximum criticality value through an atomic variable which is updated when TAOs are scheduled and completed. If the TAO is critical, a suitable core is found, and the TAO is scheduled in the corresponding work-stealing queue. 

We have implemented two strategies for finding the suitable core, one where the runtime is aware of its heterogeneity and one where it remains unaware. We chose to implement the criticality-based scheduling in these two ways in order to understand the performance effects of introducing more awareness of the resource types into the system.

For the \textit{aware} strategy, the critical TAOs are placed on a big core (randomly chosen) and non-critical TAOs are placed on a LITTLE core (randomly chosen) based on the available resources. In the \textit{unaware} strategy, whenever a critical TAO is encountered, the PTT is used to find the best performing core based on previous executions with corresponding resource width. Non-critical TAOs are then simply scheduled on a random core.  Although the overhead of the scheduling increases when searching the table, it is the most portable solution as the runtime does not need any information about the platform other than what is gathered at runtime.

\subsubsection{Weight-based Scheduling}\label{sec:bias}
The Bias scheduling scheme by Koufaty et al.~\cite{Koufaty} has been implemented by using a bias for each TAO type that describes which core is preferred. A TAO bias is estimated by analyzing the execution 
cycles from running on the respective cores and represents how suited the TAO is to the different cores. The TAO bias is dynamically calculated and aids the already existing scheduler to choose the most suitable cores. 

In our implementation we make use of the PTT to calculate a weight value for each TAO. The weight is calculated by dividing the execution time of a LITTLE core by the execution time of a big core. The calculation is performed every time a TAO is woken up in the commit-and-wakeup and the result is compared to a system wide threshold. If the weight value of a TAO is greater than the current threshold, it is an indication that this TAO type gains a larger speedup by running on a big core compared to other TAOs. Thus, if a task has a weight value greater than the threshold it is scheduled on a random big core, and if a TAO has a weight value less than the threshold, it is scheduled on a random LITTLE core. The threshold value is initially set to 1.5 and is then updated at every comparison with a weighted ratio of 1:6 to the old threshold value to represent the mean weight value of the system, i.e.~$threshold=\frac{potential weight + (old threshold value * 6)}{7}$, where the potential weight is the potential speedup calculated for the comparison.

\subsection{Task Molding}\label{sec:mold}
In addition to the scheduling extensions that focus on big/LITTLE placement, we propose an extension to resize and mold the TAOs widths. With the introduction of the PTT, we have knowledge of past executions, both which \textit{leader} core is the fastest and which resource width is the most suitable. Since the runtime already carries the notion of moldable tasks we can utilize this to make more intelligent scheduling decisions by changing the resource width. With this in mind, we implemented a molding mechanism that changes the resource width based on past execution times as well as the system load. This molding mechanism can be applied in isolation or together with the other scheduling extensions, and is also implemented within the commit-and-wakeup mechanism of XiTAO. 

We propose two policies for changing the resource width at runtime: load-based and history-based molding. The load-based molding was implemented to be able to benefit from extra resources when the system load is low. If the system load is lower than the available resources, TAOs can be molded into a larger size to better exploit the potential resources. If a DAG has a low degree of parallelism this is highly advantageous but if the system load is high the history-based molding is of more interest. The history-based molding was implemented to adjust the resource width of the TAOs to what was, in previous iterations, most suitable. Based on the core it compares trace-table entries to find the preferred resource width. Specifically, it looks within its cluster for the recorded execution times for each potential leader and resource width. The width corresponding to the best performance compared to the amount of resources, becomes the new resource width. So for a new width to be set, the recorded execution time for that width $\times$ the width has to be lower than the current execution time. This way a wider resource width is chosen only if the execution time is worth occupying more resources for. Changing the width according to history is beneficial for heterogeneous architectures where the preferred resource width of a TAO could be different between cores. Furthermore, it can also allow the TAOs to adapt to interference (e.g.~cache or memory BW oversubscription) and other types of non-architectural heterogeneities which might affect the desired resource width.

These molding policies can be used separately or together. In our implementations we use the molding policies hierarchically where the resource is adjusted to the load primarily but if the load is too high to make a justification for resizing for idle resources, then the history-based policy is used. This way, the system can both adapt to the load as well as find a good resource width targeting high performance without user intervention.


%% file: 4-eval.tex
\section{Evaluation Methodology}\label{sec:Hikey}
The evaluation of the scheduling implementations was conducted on a HiKey960 development platform equipped with an octa-core Kirin 960 processor using an ARM big.LITTLE architecture~\cite{96Boards2017}. This processor has four ARM Cortex-A73 and four Cortex-A53 cores with 3GB DDR4 SDRAM in an ARMv8 architecture node. In this processor, two L2 caches are shared among the big and the LITTLE cores, respectively. Each core has a private L1 cache.

\subsection{Evaluation Benchmarks}\label{sec:Benchmark}

We evaluate the schedulers and PTT by constructing randomly-generated DAGs with variable degree of parallelism. These irregular DAGs are composed of three types of kernels, embodied into three types of TAOs: \textit{copy}, \textit{sort} and \textit{matrix multiplication}.
We selected these three kernels as they respectively exhibit strong streaming, data-reuse and compute-bound properties. These properties cover the spectrum of behaviors usually observed in HPC kernels. 

\subsection{Kernels}\label{sec:Kernels}
When selecting the kernels, the priority was to match the desired characteristics of memory-intensive (streaming), cache-intensive (i.e. data reuse) and compute-intensive.
%
A \textit{copy} kernel handling large inputs was implemented for the streaming property. The kernel reads and writes large portions of data to memory, effectively creating a streaming behavior where the kernel has to access the main memory continuously.
For the data reuse property, a \textit{quicksort} and \textit{mergesort} kernel combination was chosen. This kernel first splits the input array into chunks and performs in-place sorting with quicksort before carrying out two levels of mergesort, effectively reusing the data within the kernel. 
%
Finally, a \textit{matrix multiplication} kernel was created for the compute-intensive property. We implemented a matrix multiplication that can benefit greatly from parallelism by ensuring that the writing of output data was done to separate cache lines for each thread while still sharing the input data.


\subsubsection*{Kernel Profiling}
To understand the behavior of the chosen kernels, we profiled them on the evaluation platform with the XiTAO runtime. 
For this purpose, the kernels were arranged into sequential chains of TAOs executing in isolation and in parallel, filling all resources.
In addition, different resource hints allowed us to see how the kernels responded to having multiple cores share the workload of a single TAO. The profiling also allowed us to see potential speedups and behaviors of the kernels on big and LITTLE cores respectively. 

For each kernel, we chose the appropriate working set size corresponding to the desired behavior. For the matrix multiplication, we chose a matrix size of 64x64 double precision elements, as it handled the largest number of computations per second. For the sorting, we chose a 262kB input array, taking up a total space of 524kB, effectively fitting in the L2 caches of the big and LITTLE cores. Finally, the copy used a 16.8MB array, taking up a total space of 33.6MB (read + write), widely exceeding the space of the L2 caches. These working set sizes also resulted in similar execution times for the kernels on the LITTLE cores. This is suitable for the benchmarks as it makes the task granularity equal across our different TAOs and it becomes simpler to monitor the fraction of the execution time for each kernel.

\begin{figure}[tbh]
\centering
\includegraphics[width=.9\columnwidth,trim=55 100 60 80, clip]{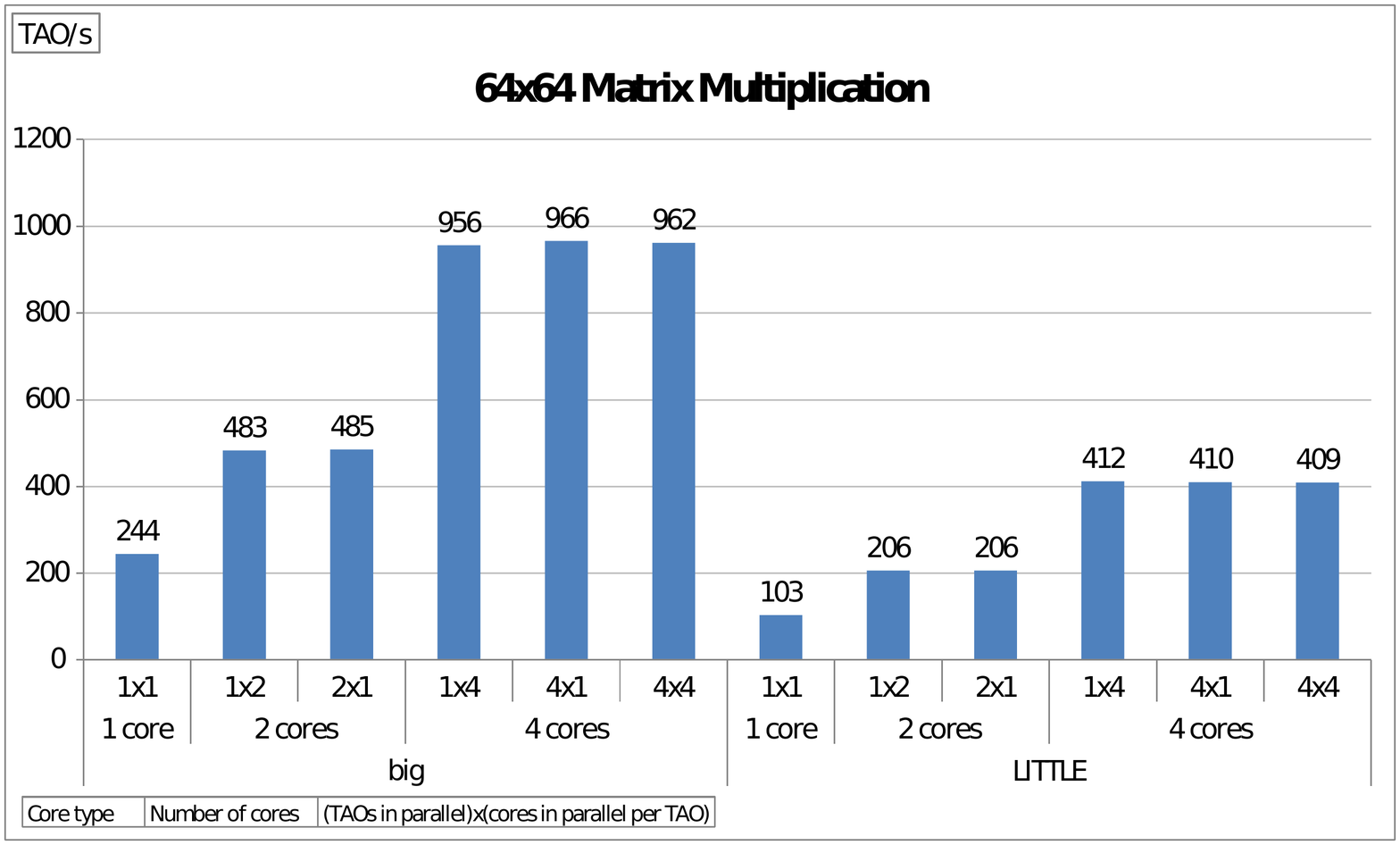}
\includegraphics[width=.9\columnwidth,trim=55 102 60 80, clip]{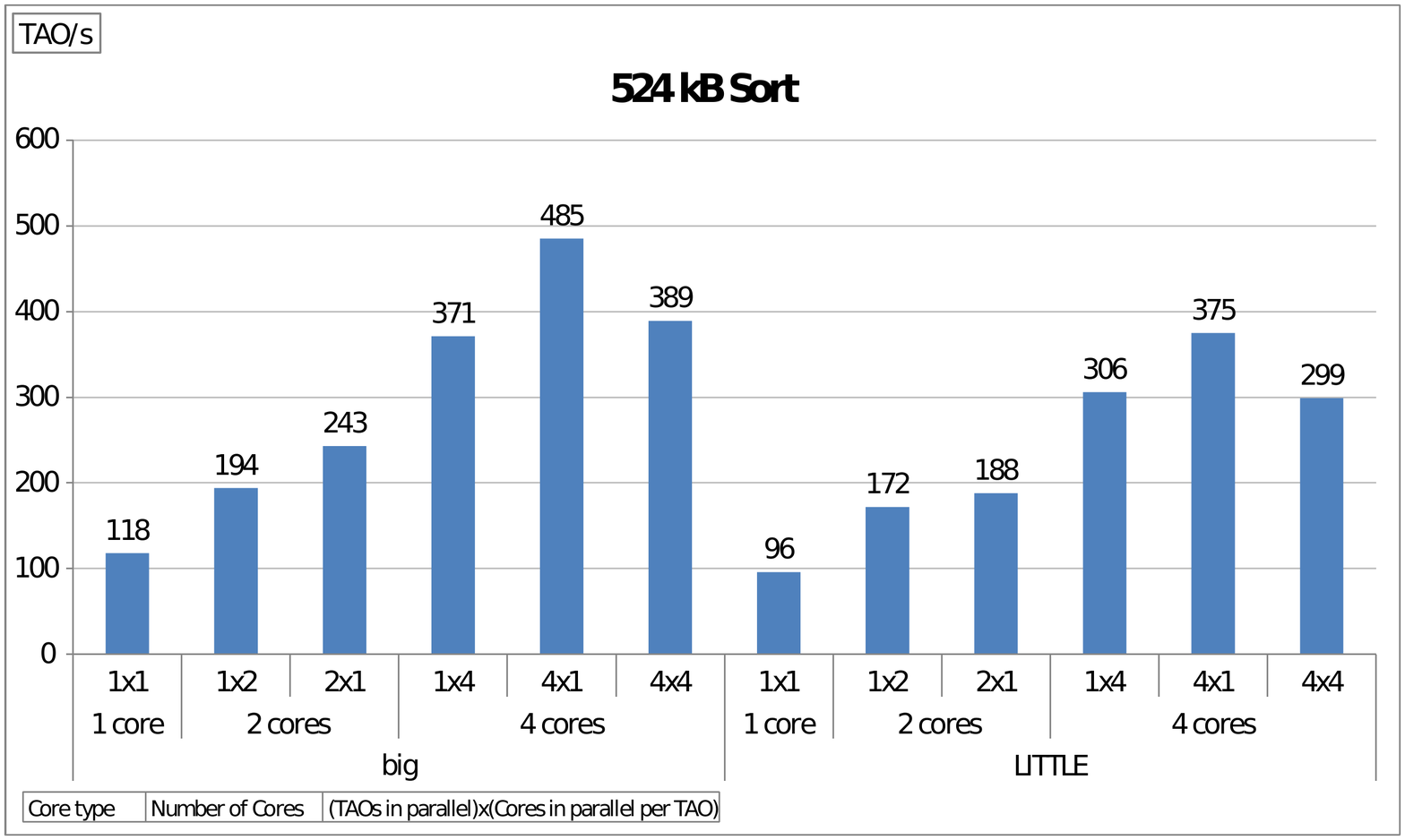}
\includegraphics[width=.9\columnwidth,trim=55 100 60 78, clip]{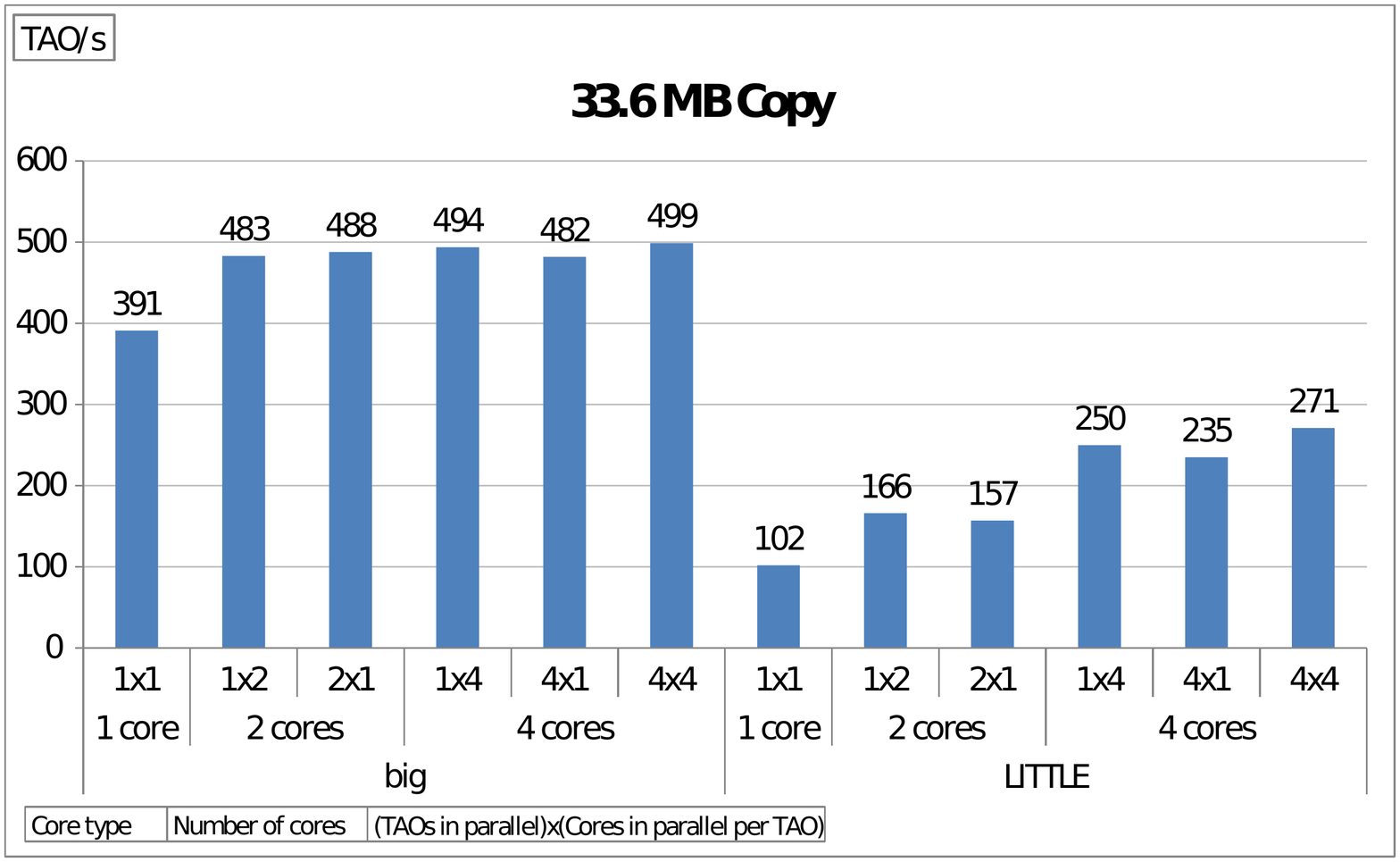}

\caption{Profiles of kernels as a function of resource width, number of concurrent TAOs and core type. 
}
\label{fig:prof}
\end{figure}

The results of the profiling can be seen in Figure~\ref{fig:prof} (top) for the matrix multiplication kernel, Figure~\ref{fig:prof} (middle) for the sorting kernel and Figure~\ref{fig:prof} (bottom) for the copy kernel. $M \times N$ denotes the number of parallel chains (\textit{M}) $\times$ the resource hint (\textit{N}).

The profiling of the matrix multiplication kernel shows a linear increase in the throughput with the number of cores, both in the case of one TAO using multiple resources and in the case of running multiple TAOs in parallel. This is expected given the compute-bound nature of the kernel. The speedup of big over LITTLE cores is 2.4$\times$ in both configurations. 

For the sorting kernel, the profiling shows a higher throughput if there are multiple TAOs rather than multiple resources working on the same TAO. This is due to the internal dependencies within this kernel. Remember that the kernel implements a mergesort reduction internally. This translates into reducing degrees of parallelism over execution stages, therefore limiting the potential performance increase for some stages, and is likely the reason for the slight decrease in performance in the larger resource hints. As allocated memory is effectively 524kB multiplied by the number of parallel chains, the performance of the 2$\times$1 and 4$\times$1 configurations suffers from interference in the shared cache since the working set no longer completely fits. Sorting only performs marginally better on the big cores rather than the LITTLE cores.

Finally, the profiling of the copy kernel shows limited performance gains from increased parallelism, especially when running on the big cores. This is the effect of reaching the memory bandwidth limit due to continuously accessing the main memory. The copy kernel performs significantly better on big cores compared to LITTLE cores, indicating that the big cores are capable of generating many more requests per unit time (saturating the memory bandwidth), while the weaker LITTLE cores are much slower and are not able to saturate the memory bandwidth.

\subsection{Randomized DAG}\label{sec:randomDAG}
To properly evaluate the performance of the scheduler, a random composition of the kernels was implemented. A random composition can effectively imitate the behavior of an irregular application that keeps changing over time.

To generate a randomized DAG 
we follow a methodology based on the DAG generator used by Topcuoglu et al. in ~\cite{Topcuoglu2002}. 
We generated three randomized DAGs with different degrees of parallelism: 1.62, 3.03 and 8.06. Each DAG consisted of 3000 TAOs, with each kernel contributing 1000 TAOs. Representations of the first 60 nodes in the DAGs are shown in Figure~\ref{fig:animals}. The figures give an idea of the structure and parallelism of these DAGs. 


\newcommand{\specialcell}[2][c]{%
  \begin{tabular}[#1]{@{}c@{}}#2\end{tabular}}

\begin{figure}[tbh]

\begin{tabular}{cc}
\includegraphics[width=0.11\columnwidth]{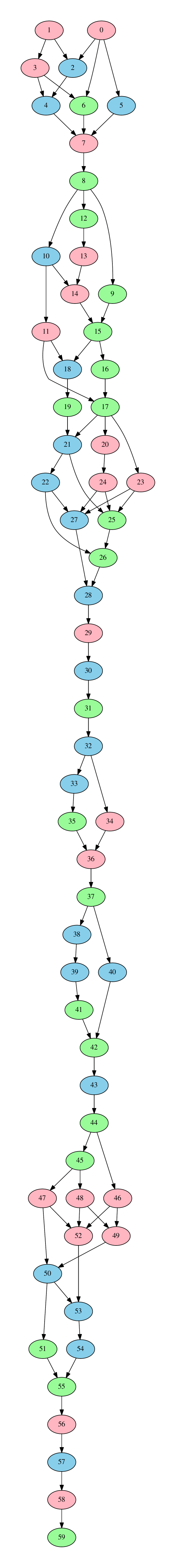} & 
\specialcell[b]{
\includegraphics[width=0.3\columnwidth]{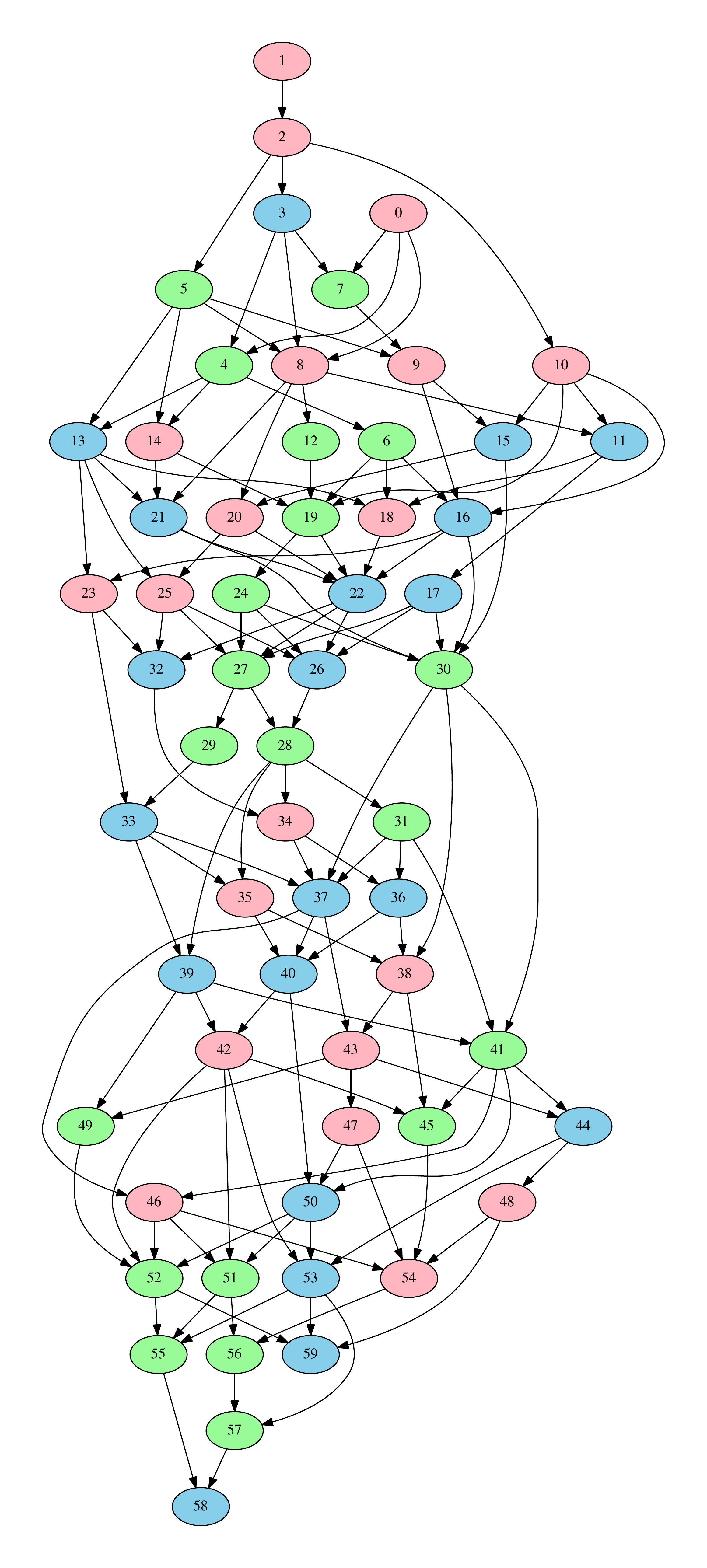} \\ \includegraphics[width=.7\columnwidth]{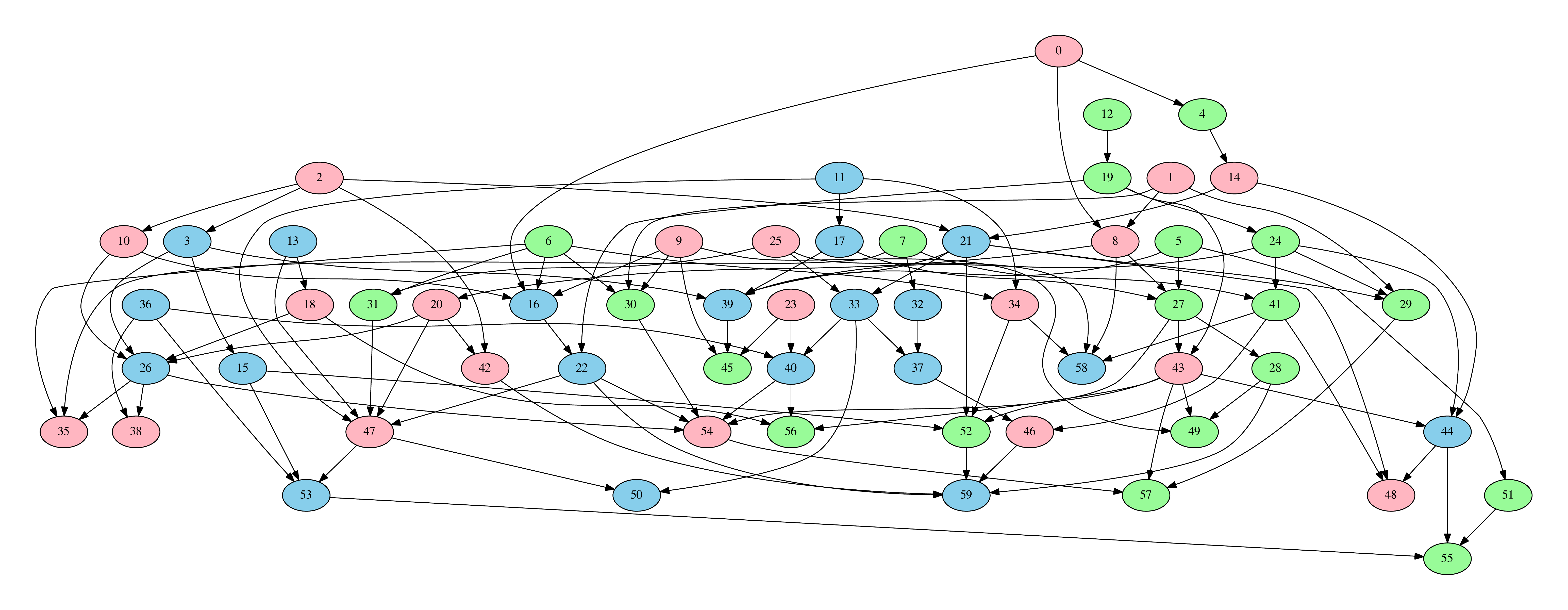}
}
\end{tabular}
    
    \caption{
    Partial versions of the randomized DAGs with degrees of parallelism 1.62 (left), 3.03 (top right) and 8.06 (bottom right). The red nodes represent matrix multiplication TAOs, the blue nodes represent sort TAOs and the green nodes represent copy TAOs.
    }\label{fig:animals}
\end{figure}

\subsection{Evaluation Metrics and Methodology}\label{sec:Evalmet}

In order to evaluate the integrated scheduling extensions we chose to compare it to the original runtime which makes programmer-directed resource-aware scheduling without considerations of heterogeneity. This allows us to effectively compare the performance impact of prioritizing TAOs based on the available heterogeneity to a case where it is not considered. The methodology also allows us to compare the usage of the PTT compared to static resource annotations.
Our evaluation metric for this is the throughput (TAOs/s) as it allows us to measure how effective the placement of the TAOs has been in terms of performance. 


The degree of parallelism for the evaluation benchmarks is another important metric to consider when analyzing the results of the scheduling implementations. We define the degree of parallelism as $\#TAOs/Cp$ where $Cp$ is the length of the critical path. 

%% file: 5-results.tex
\section{Results}\label{chap:Res}

In the following, the evaluated scheduling extensions are denoted as "Criticality-based scheduling (heterogeneity aware)" for the implementation using a predefined mapping of which cores are big and LITTLE, "Criticality-based scheduling (PTT)" for the implementation that is unaware of any heterogeneity other than what is discovered at runtime, and "Weight-based scheduling" for the implementation that selects cores based on performance gain comparisons. The molding of the resource width is used together with the Criticality-based scheduling (PTT) and the Weight-based scheduling. For the task molding, the weight-based and the history-based policies are used together. The base case in these evaluations is the runtime without any of the scheduling extensions. This runtime uses the base DPA algorithm with random work stealing. We denote this as "homogeneous scheduling". 
 
All evaluations were tested with the resource hints (widths) of all TAOs set to one and four. The working set sizes for the kernels are the profiled scenarios shown in Section~\ref{sec:Kernels}, $64\times64$ for matrix multiplication, 524kB for the sort and 33.6MB for the copy. XiTAO is configured to use 8 threads and uniform random work stealing such that each unsuccessful steal attempt is interleaved with one check of the local assembly queues (for details see~\cite{Pericas2018}).

\subsection{Randomized DAGs}\label{sec:resran}

\begin{figure}[tbh]
\centering
\begin{tabular}{c}
\includegraphics[width=0.9\columnwidth,trim=60 80 60 80, clip]{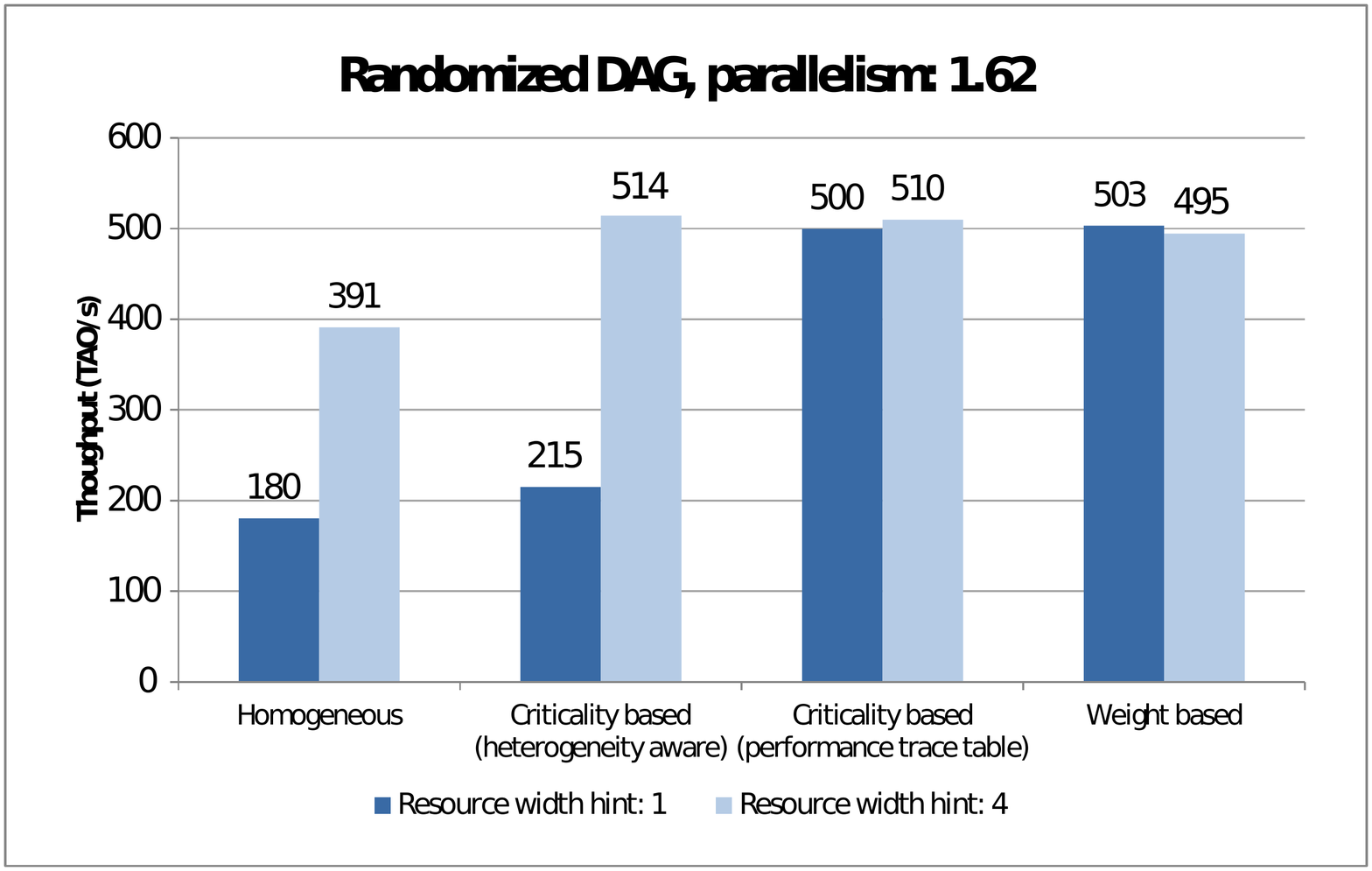} \\
\includegraphics[width=0.9\columnwidth,trim=60 265 60 310, clip]{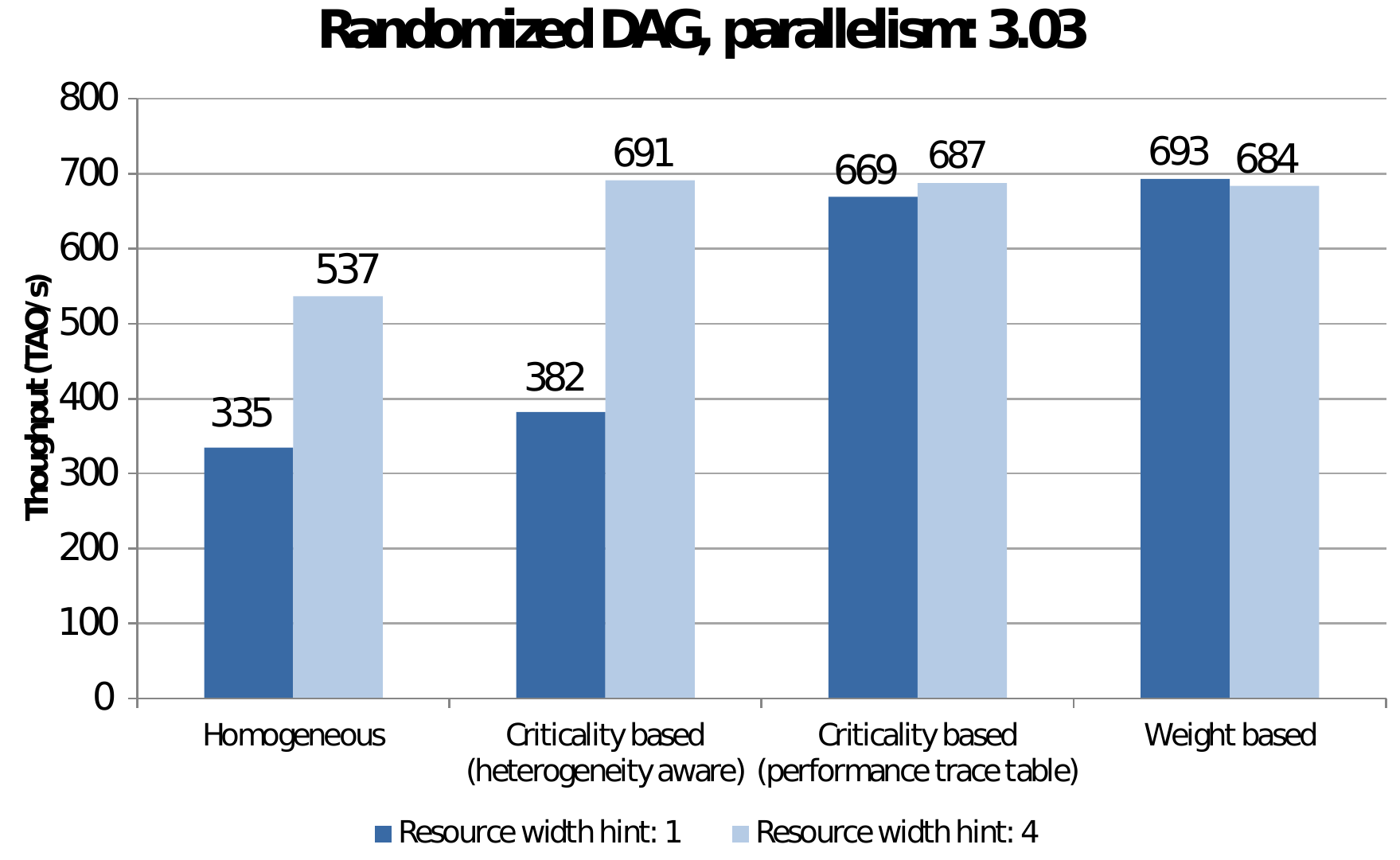} \\
\includegraphics[width=0.9\columnwidth,trim=60 265 60 310, clip]{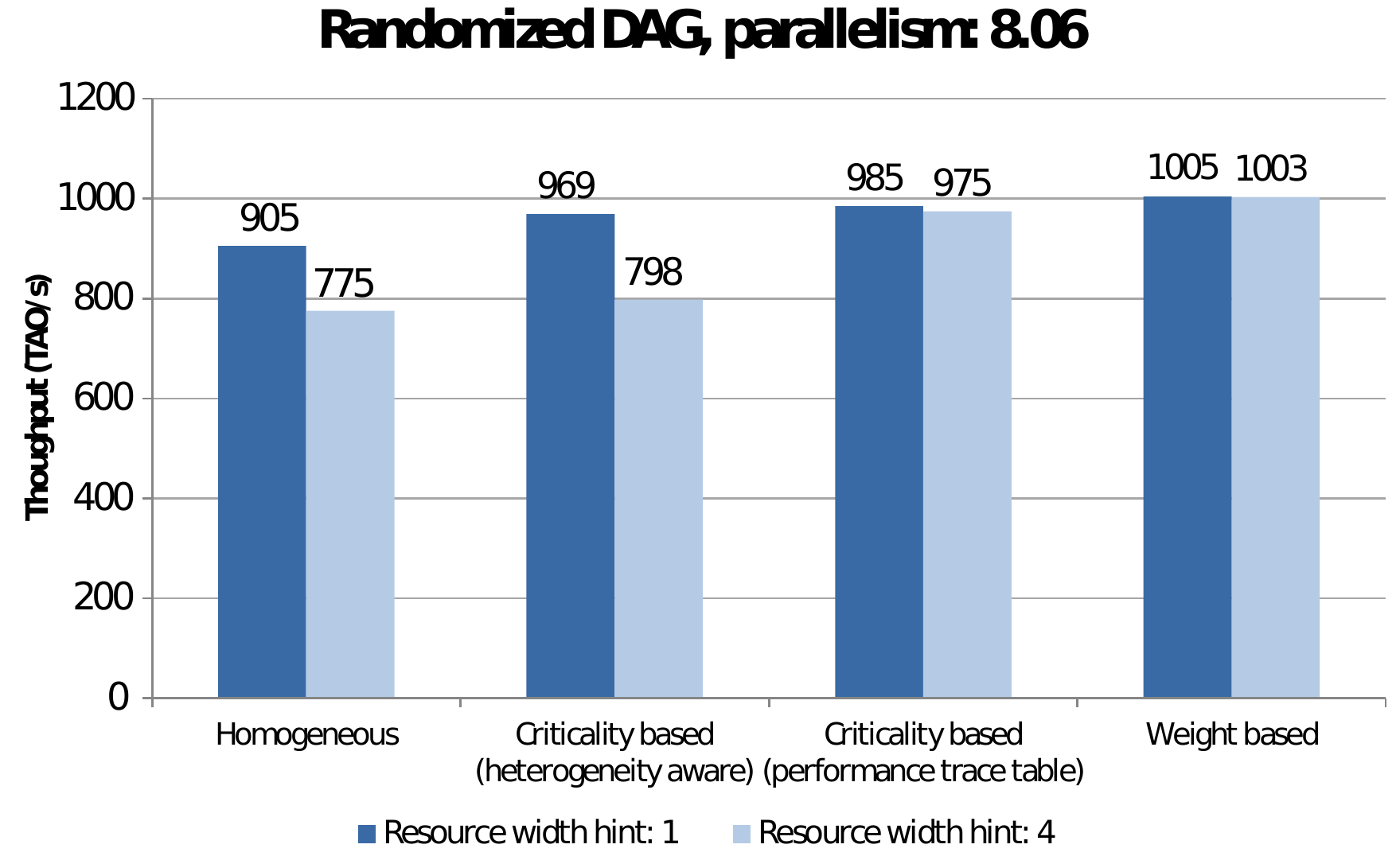}
\end{tabular}

\caption{DAGs with different degrees of parallelism}
\label{fig:result}
\end{figure}

The execution results for the DAG with a degree of parallelism of 1.62 are shown in Figure~\ref{fig:result} (top). The speedup of the heterogeneous scheduling extensions with molding (PTT) is 1.29 over the homogeneous scheduling with static resource width four and 2.78 over resource width one. For low-parallelism DAGs, using low resource with per TAO leaves most cores underutilized, as the DAG does not have enough parallelism to keep the system busy. These results show that the PTT is an effective mechanism to adapt to the degree of parallelism of the DAG of mixed-mode applications.
The heterogeneity-aware scheduler shows good performance for width four, but only 1.19 speedup over the homogeneous scheduling with width one. This shows that the benefits of criticality-aware scheduling are not enough to reap the full benefits of the platform, and that task molding is also important.

The results of the DAG with a degree of parallelism of 3.03 can be seen in Figure~\ref{fig:result} (middle). 
The approximate speedup of the scheduling extensions with molding is 1.27 over the homogeneous scheduling with resource width four and 2.03 over width one. 
The rationale is similar to the case of parallelism 1.62 Figure~\ref{fig:result} (top), but in this case, because of the higher degree of parallelism, the width-1 schedulers are able to achieve higher utilization, leading to smaller speed-ups for the PTT and width-4 schedulers. 
For the heterogeneity-aware criticality scheduling, width-4 shows a similar speedup with 1.28 over the corresponding homogeneous scheduling base case.
The heterogeneity-aware scheduling with width-1 achieves a speed-up of 1.14 over homogeneous scheduling with width-1, showing how the effects of criticality-based scheduling only, are slightly smaller in this case. This is to be expected, as DAGs with higher degrees of parallelism are less sensitive to the critical path. 

Finally, the results of the DAG with a degree of parallelism of 8.06 can be seen in Figure~\ref{fig:result} (bottom). For this benchmark, resource width one shows the best throughput for the homogeneous scheduling. 
The reason is that now even the width-1 static schedulers can keep all the cores busy. The width-4 schedulers, on the other hand, suffer a slight penalty due to the fact that not all TAOs scale linearly with the resources. In particular, the sort TAOs do not keep all resources busy, as they internally implement a mergesort reduction. 
The speedup of the scheduling extension with molding is 1.1 over the homogeneous scheduling with width one and 1.28 over width four. This shows that despite the higher parallelism using the PTT is still useful in this case. In this case, the benefit comes from detecting cases in which scheduling too many TAOs in parallel leads to excessive resource interference. The PTT can detect such cases and dynamically reduce TAO parallelism in order to limit interference.

\subsection{Task Molding Evaluation}\label{sec:molding_eval}
In Table~\ref{tab:moldover_bias} and Table~\ref{tab:moldover_crit} the throughput values for the three randomized DAGs are presented with the two respective scheduling extensions, criticality-based scheduling (PTT) and weight-based scheduling, with and without molding. 
In these tables, the resource hints corresponds to the resource hints resulting in highest throughput for the base case homogeneous scheduling.
Small overheads of using molding together with the two scheduling extensions can be seen in the case of 1.62 or 3.03 degree of parallelism, where at most the weight-based scheduling without molding achieves a 1.01 speedup over the corresponding case with molding. In the heavily parallel DAG the molding of the resources instead show a performance gain over the case without molding, at most the speedup is 1.08 with the criticality-based scheduling. This indicates that even for highly parallel applications there are cases in which saturating the full chip resources results in a performance decrease. Instead, by dynamically learning the platform's behavior, the PTT can limit such cases of resource oversubscription and achieve higher performance than a fully greedy scheduler.

\begin{table}[tbh]
\centering
\caption{Impact of task molding on the weight-based scheduling.
}
\label{tab:moldover_bias}
\begin{tabular}{|lll|}
\hline
\textbf{DAG(Degree of Parallelism)} &Without Molding &  With Molding \\
\hline
1.62 (Resource Hint = 4)            & 497          & 495\\
3.03 (Resource Hint = 4)            & 693          & 684\\
8.06 (Resource Hint = 1)            & 946          & 1005\\                                                       \hline              
\end{tabular}
\end{table}

\begin{table}[tbh]
\centering
\caption{Impact of task molding on the criticality-based scheduling (performance trace table).
}
\label{tab:moldover_crit}
\begin{tabular}{|lll|}
\hline
\textbf{DAG(Degree of Parallelism)} &  Without Molding &     With Molding \\
\hline
1.62 (Resource Hint = 4)                       & 510                                         & 510                                                   \\
3.03 (Resource Hint = 4)                       & 695                                         & 687                                                   \\
8.06 (Resource Hint = 1)                     & 909                                         & 985\\
\hline
\end{tabular}
\end{table}

%% file: 6-conc.tex
\section{Conclusion}\label{chap:Conc}

In this paper we present different scheduling extensions for the scheduling of mixed-mode parallelism on heterogeneous platforms. Specifically, we extend the XiTAO's homogeneous scheduler to consider heterogeneity. Techniques from related heterogeneous scheduling methods together with moldable tasks from XiTAO were used in several experimental scheduling configurations. The extensions target a single ISA heterogeneous multicore architecture and are evaluated on an ARM big.LITTLE architecture for performance increases. 
In addition, a performance trace table is introduced to record execution times of tasks on the different cores with different widths. 


The evaluation shows the benefit of feedback-directed resource partitioning and heterogeneity-aware scheduling. In scenarios where a programmer made a sub-optimal choice of TAO parallelism, we observe improvements of 28-182\% depending on the parallelism degree of the DAG. The task molding extension, supported by the newly introduced performance trace table (PTT), is able to find an appropriate resource width efficiently. In scenarios in which an appropriate resource width was statically chosen, we observe improvements that range from 10\% up to 45\% for both the schedulers that target criticality and the scheduler that targets the performance trade-offs. 

\section*{Acknowledgment}

The research leading to these results has received funding from the European Union's Horizon 2020 Programme under the LEGaTO Project (www.legato-project.eu), grant agreement number 780681.